\documentclass[12pt]{article}
\usepackage{amsmath}
\usepackage{epsfig}
\usepackage[tableposition=top]{caption}
\Roman{section}
\renewcommand{\thesection}{\arabic{section}}

\newcommand{\eq}[1]{\begin{equation}#1\end{equation}}
\newcommand{\naw}[1]{\left(#1\right)}

\newcommand{\com}[1]{\left[#1\right]}

\newcommand{\poisson}[1]{\left\{#1\right\}}

\begin{document}

\begin{center}
\textsc{\Large{Note on log-periodic description of 2008 financial crash}}

\emph{Katarzyna Bolonek-Lason\footnote{Department of Statistical Methods, Faculty of Economics and Sociology, University of Lodz, Poland. The work paper funded by resources on science period 2008-2010, Nr N N111 306335 }},
\emph{Piotr Kosinski\footnote{Department of Theoretical Physics II, Faculty of Physics and Informatics University of Lodz, Poland}}
\end{center}

We analyze the financial crash in 2008 for different financial markets from the point of view of log-periodic function model. In particular, we consider Dow Jones index, DAX index and Hang Seng index. We shortly discuss the possible relation of the theory of critical phenomena in physics to financial markets.    

\section{Introduction}

\quad In recent years many papers have appeared which deal with applications of the methods of statistical physics to the stock market description, in particular - to the analysis and prediction of financial crashes.

\quad  The interactions between physical particles are usually short-ranged. In spite of that physical systems exhibit, under certain conditions, cooperative behaviour. The Onsager solution \cite{Onsager} of the twodimensional Ising model may be viewed as the first mathematically precise derivation of this fact from first principles.

\qquad It is very tempting to pursue this idea when considering the stock markets dynamics; here also a cooperative phenomena appear, for example in  the form of financial crashes which may arise if the players in bulk decide to sell their shares. Moreover, they are usually neither associated with specific new items nor caused by the fact that a large group of agents communicate with each other. On the contrary, the traders can imitate the opinions of rather restricted set of nearest neighbors ("short-range interactions"). The mechanism leading to the observed stock market behaviour has been proposed  by Johansen et al \cite{Sorn-Johan}$\div $\cite{Sorn}. They assume that the traders influence each other only locally and either imitate the opinions of their nearest neighbors or take decisions independently. In the former case, if the tendency for traders to imitate their neighbors increases up  to a certain critical point, many traders may place the same order - sell at the same time thus causing a crash. In the latter the buyers and sellers disagree with each other and roughly balance out. These ideas can be formulated analytically providing quantitative relationships between some relevant parameters like critical exponents and log-periodic frequencies (see Sec. 2).

\qquad The resulting model has been applied to many financial crashes (including real-estate \cite{Zhou}, oil \cite{Sorn-Wood} bubbles as well as some Chinese stock market bubbles \cite{Jiang})and still invokes controversies  \cite{laloux}, \cite{bree}. The main one concerns, of course, the possibility of prediction when the crash will actually occur. According to the model the crash time is the accumulation point of geometric series of local minima. However, the very recognition of proper minima to be taken into account is not always easy.
Even if we find it possible the result may appear unsatisfactory. As an example consider the time series of Hang Seng index in period 2003-2008. It has four minima at the distances: $t_2-t_1=512$, $t_3-t_2=191$, $t_4-t_3=102$. The value of scaling factor $\lambda$, as defined by eq. (\ref{e}) below, reads $\lambda=2.68$ or $\lambda=1.87$, depending on the points selected.

\qquad It is, therefore, an open question whether the analogy between the critical phenomena in physics and the behaviour of stock market is deep or rather superficial. We remain slightly sceptical. However, we believe that further analysis is necessary. In the present paper the analysis of financial crash in 2008 for different markets is presented.\\
The paper is organized as follows. In Sec. 2 the theoretical model introduced by Johansen et al. is sketched. Sec. 3 contains empirical results. Last section is devoted to some conclusions.

\begin{figure}[htbp]
\centering
\includegraphics[width=0.7\textwidth]{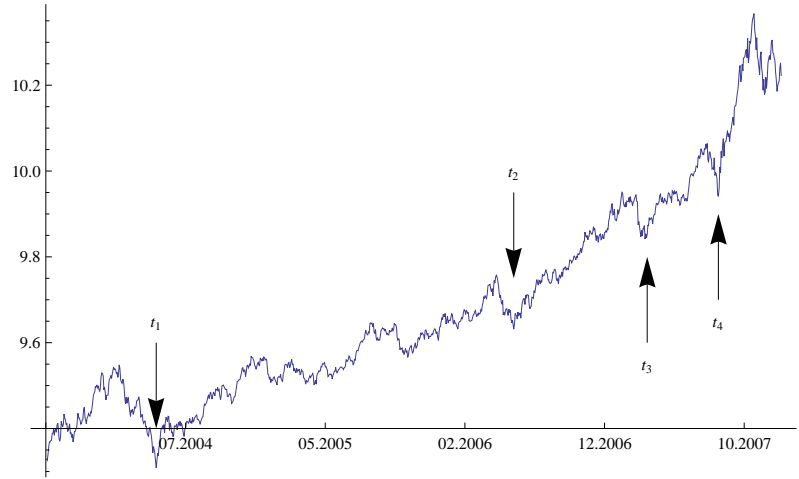}
\caption{Logarithm of Hang Seng index over the period 2003-2007.}
\label{rys6}
\end{figure}

\section{The JLS model}
The following theoretical model has been introduced by Johansen et al. \cite{Sorn-Johan}$\div $\cite{Sorn}. One considers the hazard rate which is the probability per unit time that the crash will happen in the next instant if it has not happened yet. Therefore, the hazard rate is defined as: 
\eq{h\naw{t}=\frac{q\naw{t}}{1-Q\naw{t}},}
where $q\naw{t}$ is the probability density function of the time of the crash, $Q\naw{t}$ - the cumulative distribution function. The dynamics of the asset price is given by:
\eq{dp=\mu\naw{t}p\naw{t}dt-\kappa p\naw{t}dj.\label{1}}    
The authors assume that, in the case of a crash, the price drops by a fixed percentage $\kappa\in\naw{0,1}$; \emph{j} denote a jump process whose value is zero before the crash and one afterwards; $\mu\naw{t}$ is chosen so that the price process satisfies the condition:
\eq{\forall t'>t \qquad E_t\com{p\naw{t'}}=p\naw{t}.\label{a}}
This condition yields
\eq{\mu\naw{t}=\kappa h\naw{t}.\label{b}} 
Inserting eq.(\ref{b}) into eq. (\ref{1}), we have
\eq{log\com{\frac{p\naw{t}}{p\naw{t_0}}}=\kappa\int^{t}_{t_0}h\naw{t'}dt' \label{cc}}
before the crash.
In order to be able to use eq. (\ref{cc}) the information concerning the behaviour of hazard rate $h\naw{t}$ is needed. 
Johansen et al. \cite{Joh-Led-Sorn} proposed the form of $h\naw{t}$ based on the above - mentioned analogy with statistical physics. Namely, they assumed that any agent $i$ can be only in one of two possible states $s_i\in\poisson{-1,+1}$ (buy/sell). The state of \emph{i}-th agent is determined by:
\eq{s_i=sign\naw{K\sum_{j\in N\naw{i}}s_j+\sigma\varepsilon_i}}
where the $sign\naw{\cdot}$ function is equal to $+1$ ($-1$) for positive (negative) numbers. $N\naw{i}$ denotes the set of neighbours of \emph{i}-th agent; only their decisions can influence him/her. \emph{K} is positive constant, $\varepsilon_i$ are independently distributed according to the standard normal distribution. \emph{K} is called coupling strength and control the tendency towards imitation; the tendency towards idiosyncratic behavior is governed by $\sigma$.\\
The model formulated in the above way resembles the Ising model solved (in twodimensional case) by Onsager. Thus one can expect that there exists a critical value of $K$, $K_c$, such that the system becomes very sensitive to small global perturbations (in the physical language, it gets magnetized after applying an arbitrarily small external magnetic field and exhibits large-scale fluctuations). It is assumed that the hazard rate which measures the reaction on small perturbations behaves as magnetic susceptibility i.e. 
\eq{h\naw{t}\approx B_0\naw{t_c-t}^{-\beta}+B_1\naw{t_c-t}^{-\beta}\cos\com{\omega\log\naw{t_c-t}+\psi'}.\label{c}}
where the critical time $t_c$ is defined as the first time such that $K\naw{t_c}=K_c$\\
Putting eq.(\ref{c}) into eq.(\ref{cc}), one arrives at the following formula for the evolution of the price before the crash
\eq{\log\com{p\naw{t}}\approx A+B\naw{t_c-t}^\alpha \poisson{1+C\cos\com{\omega\log\naw{t_c-t}+\phi}}\label{d}}
where $\phi$ is another phase constant.

\section{Empirical Results}

\subsection{Fitting the log-periodic function to empirical date}
The empirical search for log-periodic function describing the finacial market was carried out in many papers \cite{bartolozzi}-\cite{cajueiro}. In order to find seven parameters of log-periodic function we must minimize the following expression:
\eq{\sum^{t_n}_{t=t_1}\naw{y_t-\hat{y}_t}^2=\sum_{t=t_1}^{t_n}\poisson{y_t-A-B\naw{t_c-t}^\alpha\naw{1+C\cos\naw{\omega\log\naw{t_c-t}+\phi}}}^2,\label{dd}}         
where $y_t$ is the logarithm of real index price, $\hat{y}_t$ is the date point as predicted by the model, and $t_n<t_c$. However, the minimalisation is not an easy task. The noisy data and a fitting function with large number of degrees of freedom cause many local minima of eq. (\ref{dd}) to exist where the minimalisation algorithm can get trapped. In some papers \cite{Joh-Led-Sorn},\cite{cajueiro} the parameters $A$, $B$, $C$ are reduced by requiring first that eq.(\ref{dd}) has at the minimum the vanishing derivatives with respect to $A$, $B$, $C$. In this way one gets 3 linear equation for the latter. Solving them allows to express A, B and C as functions of remaining parameters. However, the form of these functions is very complicated and, moreover, one has still quite a lot of parameters controlling the fit.\\
For this reason we look for the best fit in the following way. The log-periodic corrections to scaling imply the existence of hierarchy of characteristic scales in space or time. The linear scale is determined by three consecutive minima or maxima of log-periodic oscillations \cite{drozdz1},\cite{sorn2}       
\eq{\lambda =\frac{t_{n+1}-t_n}{t_{n+2}-t_{n+1}}.\label{e}}
In eq. (\ref{d}) the linear scale is contained in parameter $\omega = 2\pi / \ln \lambda$. The authors of papers \cite{bartolozzi}-\cite{drozdz2} 
postulated that $\lambda$ is the universal parameter approximately equal 2. However, eq. (\ref{e}) shows that the $\lambda$ parameter may in principle depend on the specific geometry and structure of the system.

\quad  Assuming that the spacing beetwen successive values of $t_n$ approaches zero as $n$ becomes large and $t_n$ converges to $t_c$ (more precisely, assuming eq. (\ref{e}) to hold with $\lambda >1$), we can show that:
\eq{t_c=\frac{t^2_{n+1}-t_{n+2}t_n}{2t_{n+1}-t_n-t_{n+2}}.\label{ee}}   
Thus we determine $\lambda$ and $t_c$ from three successive observed values of $t_n$. There remain five parameters which are then determined with the help of nonlinear regression (NonlinearRegression packet of Mathematica). This preliminary fit allows us to put some constrains on parameters we are looking for. Using these constraints we minimize the right hand side of eq. (\ref{dd}) with respect to all seven parameters; due to the constraints obtained in the first step it is now unlikely to get false minima.

\subsection{Dow Jones Industrial Average Index}
We analyze the bubble data for the period from 02-Feb-2003 to 31-Oct-2007. One must keep in mind that the system is scale invariant only nearby the critical point. For this reason we carry out the procedure of fitting the log-periodic function to real data only for data between ten and twenty months before the crash. From the period 18.01.2006 - 29.08.2007 three successive minima have been selected in order to calulate $\lambda$ and $t_c$. From equations (\ref{e}) and (\ref{ee}) we find $\lambda =1.4$, and $t_c = 18.09.2008$ (which corresponds to the point 672).  We looked for the parameters of the best fit of the log-periodic function:
\eq{\log\com{p\naw{t}}\approx A+B\naw{672-t}^\alpha \poisson{1+C\cos\com{18.66*\log\naw{672-t}+\phi}}}
to real data. The values of these parameters are presented in Table \ref{T1}. 
 The value of estimated variance, defined as $MSE=\naw{\sum^{t_n}_{t=t_1}\naw{y_t-\hat{y}_t}^2}/df$, is equal to $3.68\times 10^{-4}$;                 here $df=402$ is the number of degrees of freedom.    
\begin{table}
\caption{The values of estimated parameters of log-periodic function for Dow Jones index from 18.01.2006 - 29.08.2007.}
\begin{tabular}{c|c|c|c}
 & \text{Estimate} & \text{Asymptotic SE} & \text{Confidence Interval} \\ \hline
 A & 8.68961 & 0.0659957 & \{8.55987,8.81935\} \\
 B & -0.00405699 & 0.00410804 & \{-0.0121329,0.00401894\} \\
 C & 0.0565023 & 0.0114346 & \{0.0340232,0.0789814\} \\
 $\alpha$  & 0.737919 & 0.13599 & \{0.470578,1.00526\} \\
 $\phi$  & 2.45213 & 0.0632643 & \{2.32776,2.5765\} 
\end{tabular}
\label{T1}
\end{table}

\begin{figure}[htbp]
\centering
\includegraphics[width=0.7\textwidth]{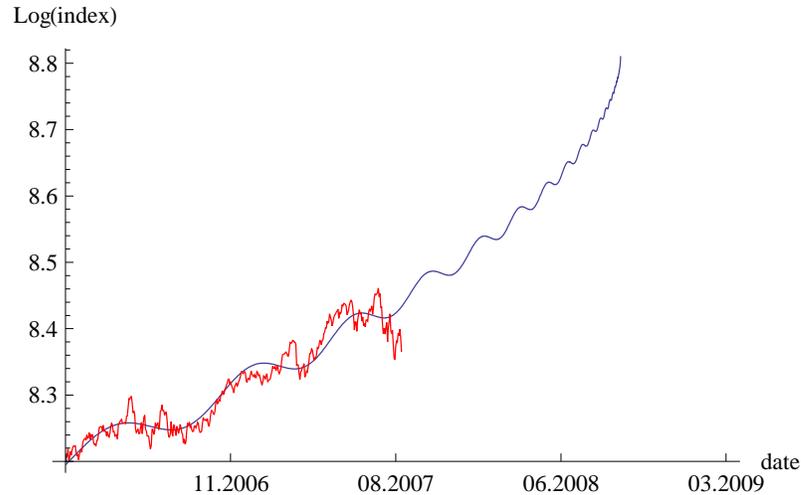}
\caption{Logarithm of Dow Jones index over the period 18.01.2006 - 29.08.2007 versus its corresponding log-periodic representation.}
\label{rys1}
\end{figure}

\begin{figure}[htbp]
\centering
\includegraphics[width=0.7\textwidth]{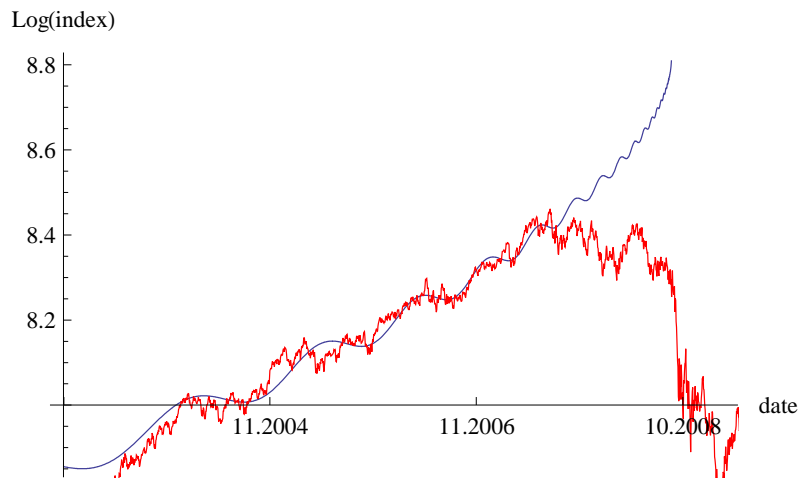}
\caption{Logarithm of Dow Jones index over the period 12.11.2002 - 26.10.2009 versus its corresponding log-periodic representation.}
\label{rys2}
\end{figure}

The parameters estimated from nonlinear regression do not allow to determine very restrictive constraints for all parametrs of eq. (\ref{d}). We assume that $15<\omega <20$, $B<0$, $0<\phi <2\pi$, $0.1<\alpha <1$, $06.06.2008<t_c<28.10.2008$. The ranges of acceptable values for parameters are greater than the confidence intervals (Table 1) because we haven't estimated the values of $\lambda$ and $t_c$. With the above constraints the function (\ref{dd}) was minimalized. We obtained the following values of the parameters: $A\rightarrow 8.8106$, $B\rightarrow -0.0165957$, $C\rightarrow -0.0444881$, $\alpha\rightarrow 0.554188$, $t_c \rightarrow 672.319$, $\varphi\rightarrow 0$, $\omega\rightarrow 19.5637$,
  and the estimated variance is $3.51\times 10^{-4}$. The determination coefficient equals 93$\%$ and is a slightly greater than for the case when $\lambda$ and $t_c$ were calculated from eq. (\ref{e}) and (\ref{ee}).       
It is very interesting that the estimated date of crash is the same as the one which has been calulated from eq. (\ref{ee}). Figure \ref{rys2} shows that the estimated log-periodic function describes very well the price dynamics in the whole period under consideration, despite of the fact that it was found by considering a short period (Figure \ref{rys1}).  

\subsection{DAX Index} 
\begin{figure}[htbp]
\centering
\includegraphics[width=0.7\textwidth]{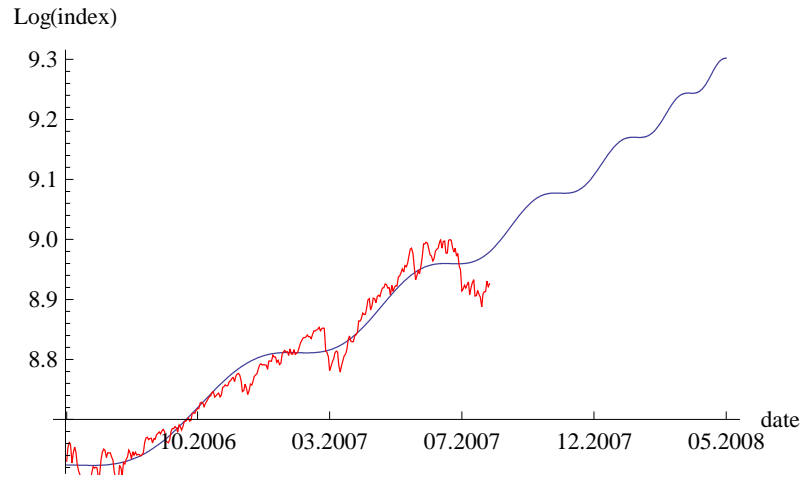}
\caption{Logarithm of DAX index over the period 25.05.2006 - 27.08.2007 versus its corresponding log-periodic representation.}
\label{rys3}
\end{figure}

DAX index  measures the development of the 30 largest and best-performing companies on the German equities market and represents around 80\% of the market capital authorized in Germany. 
From the data concerning the period 25.05.2006-27.08.2007 one finds using eqs. (\ref{e}), (\ref{ee}) $\lambda=1.55$ and $t_c=26.05.2008$. Other estimated parameters are shown in Table \ref{T2}. The estimated variance is equal $6.44\times 10^{-4}$.
\begin{table}
\caption{The values of estimated parameters of log-periodic function for DAX index from 25.05.2006 - 27.08.2007.}
\begin{tabular}{c|c|c|c}
 & \text{Estimate} & \text{Asymptotic SE} & \text{Confidence Interval} \\ \hline
A & 9.4309 & 0.0988434 & \{9.23643,9.62538\} \\
 B & -0.0158149 & 0.0118548 & \{-0.0391392,0.00750939\} \\
C & 0.0440315 & 0.00673482 & \{0.0307807,0.0572823\} \\
$ \alpha$  & 0.632698 & 0.101428 & \{0.433139,0.832257\} \\
 $\varphi$  & 0.245255 & 0.074815 & \{0.0980567,0.392454\}
\end{tabular}
\label{T2}
\end{table} 
On the basis of these values, we assume the following constraints: $03.03.2008<t_c <30.12.2008$, $10<\omega<18$, $B<0$, $0.1<\alpha <1$. The minimalization of eq. (\ref{dd}) gave the following results: $A\rightarrow 9.52346$, $B\rightarrow -0.0130455$, $C\rightarrow -0.0390518$, $\alpha\rightarrow 0.669289$, $\tau \rightarrow 570.63\approx 21.08.2008 $ (actually, the crash occured in October; however, the DAX index value fell by above 17\% already in the period 08.-10.2008), $\varphi \rightarrow -2.33041$, $\omega \rightarrow 18$. The estimated variance is equal to $6.026\times 10^{-4}$ which means that the regression curve approximates the real data in 96\%.    In contrast to Dow Jones the estimated log-periodic function of DAX does not describe very well the price dynamics in the whole period (Figure \ref{rys3}).  
\begin{figure}[htbp]
\centering
\includegraphics[width=0.7\textwidth]{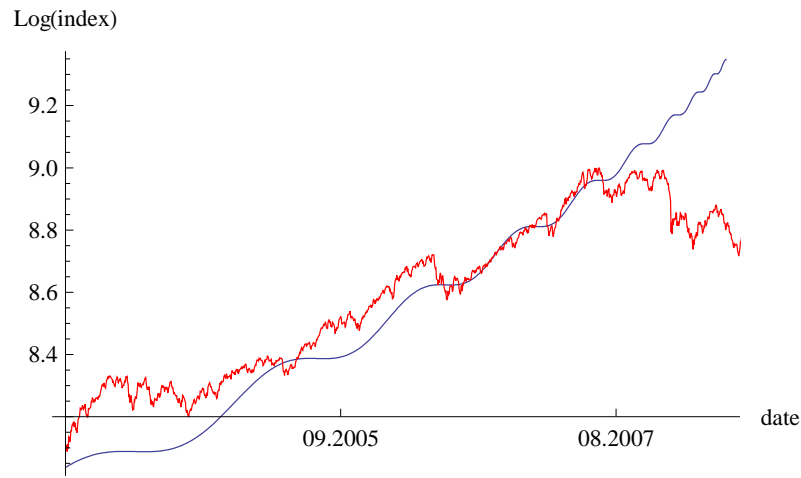}
\caption{Logarithm of DAX index over the period 29.09.2003 - 24.09.2009 versus its corresponding log-periodic representation.}
\label{rys4}
\end{figure}  
It reflects, however, the decrease of price in August 2004 and October 2006.

\subsection{Hang Seng Index}
The Hang Seng Index is used to record and monitor daily changes of the largest companies of the Hong Kong stock market and is the main indicator of the overall market performance in Hong Kong.\\
In this case we decided to check whether it is possible to get a reasonable description by taking into account the data from a longer period, not from the one directly preceding the crash. In statistical physics the universal behaviour is expected in the neighbourhood of critical point while far from it the properties of the system may be very specific and depend on particular dynamics. However, it appears here (see below) that the fit with the help of log-periodic function is still satisfactory. 

\begin{figure}[htbp]
\centering
\includegraphics[width=0.7\textwidth]{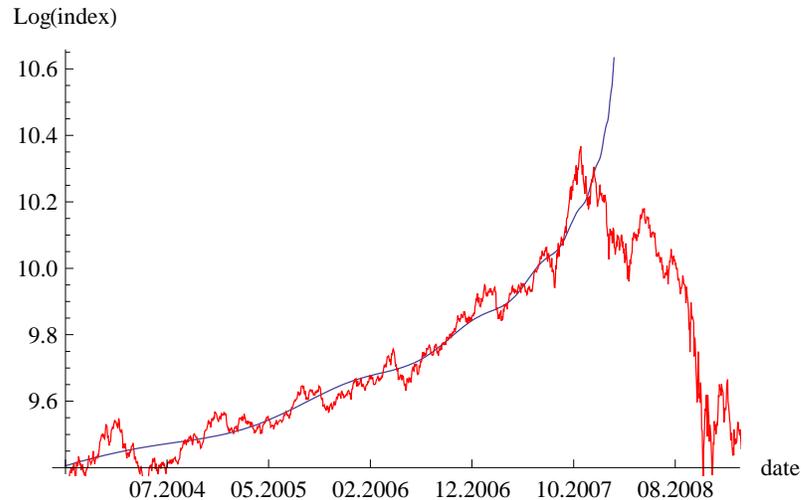}
\caption{Logarithm of Hang Seng index over the period 29.09.2003 - 09.09.2009 versus its corresponding log-periodic representation.}
\label{rys5}
\end{figure}

\quad We take four successive minima from which, from eq. (\ref{e}) and (\ref{ee}), we get the average value of parameter $\lambda\approx 2.3$ and $t_c=1087\rightarrow 18.02.2008$. Using them, we put the constraints: $990<t_c<1200$, $7<\omega<11$, $0<\phi<2\pi$, $B<0$, $0.1<\alpha<1$. The parameters minimalizing eq. (\ref{dd}) read: $A\rightarrow 11.3902$, $B \rightarrow -0.550977$, $C \rightarrow -0.00857998$, $\alpha\rightarrow 0.184217$, $t_c\rightarrow 1085.13$, $\phi\rightarrow 0$, $\omega\rightarrow 8.90061$. The regression function approximates the real data points as much as in 96\% (Figure \ref{rys5}); the system had 972 degrees of freedom.  

\subsection{The next terms of log-periodic function}
So far we were considering the log-periodic function only keeping the first term of Fourier expansion in eq. (\ref{d}). Now, we analyze the price dynamics of Dow Jones index taking into account both the first and second term of Fourier expansion. We choose the same time period as subsection 3.2. The log-periodic function is described by:
\eq{\log\com{p\naw{t}}\approx A+B\naw{t_c-t}^\alpha \poisson{1+C\cos\com{\omega\log\naw{t_c-t}+\phi}+D\cos\com{2\omega\log\naw{t_c-t}+\psi}}.}    
With the use of the same constraints on parameters as in subsection 3.2, we have: $A\rightarrow 8.93833$, $B\rightarrow -0.0423764$, $C\rightarrow -0.0367758$, $D\rightarrow -0.0173993$, $\alpha \rightarrow 0.439734$,$t_c \rightarrow 667.641\approx 12.09.2008$, $\psi \rightarrow -8.64053$, $\phi
\rightarrow 0$, $\omega \rightarrow 19.6062$. The regression curve approximates the real data in 94,2\% which, as expected, is more than in the case when only the first term is taken into account. The approximated function describes the price dynamics of bubbles better but the improvement is not dramatic (Figure \ref{rys4}). We could expect that the next terms of Fourier expansion of log-periodic function will increase the determination coefficient but not significantly.   
\begin{figure}[htbp]
\centering
\includegraphics[width=0.7\textwidth]{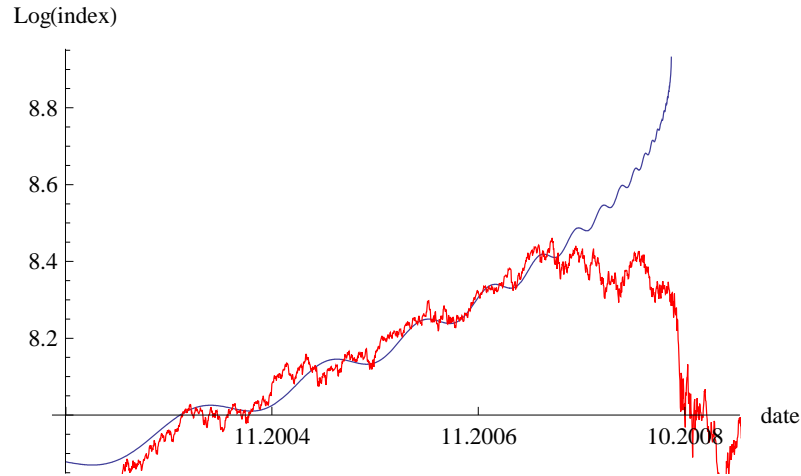}
\caption{Logarithm of Dow Jones index over the period 12.11.2002 - 26.10.2009 versus its corresponding log-periodic (first and second terms of Fourier expansion) representation.}
\label{rys4}
\end{figure}

\newpage

\section{Conclusions}
We analyzed the financial crash in October 2008 on the basis of indices for different markets. Only in the case of Hang Seng index we received the value of critical time compatible with the definition of critical point as the point where the central trend of price dynamics changes its direction \cite{drozdz2}.  In other cases it was rather the moment when the index price drops sharply in short period.
This finding supports the scepticism concerning the possibility of exact prediction of crash date on the basis of log-periodic function approach expressed already in Refs. \cite{laloux}, \cite{bree}, \cite{feigenbaum1}.

\qquad The same financial crashes were discussed in the papers \cite{bree} and \cite{drozdz2}. Their authors obtained $\lambda=2$ and $\lambda=3$, respectively. The differences seem to result from the way the minima of time series were selected. Therefore, we think that the predictive power of the approach based on log-periodic functions is doubtful. However, the aposteriori description of the price history is quite satisfactory as it is seen from the large value of determination coefficient.     

\qquad It is very tempting to compare the dynamics of market indices with the critical phenomena in condensed matter physics. The latter offer the well understood example of the situation when local interactions lead to large-scale cooperative behaviour; the main idea is the emergence of scaling invariance, the main tool - the renormalization group transformations which provide a natural framework for studying the universality phenomena. However, it is rather unlikely that the explanation offered by the theory of critical phenomena is sufficiently general to cover all cases of emergence of cooperative behaviour in complex dynamical systems.                                                           

\quad The analogy with statistical mechanics can be also questioned due to the fact that, contrary to the case of condensed matter systems, the description based of scaling seems to work reasonably also quite far from critical points. As it had been shown above, in the case of Hang Seng index the determination coefficient was equal 96\%. Guided by this value we have verified also the quality of approximation for the data from longer period 2003-2008 for remaining indices. The relevant coefficients for Dow Jones and DAX were equal 97.47\% and 97\%, respectively. Therefore, the distance from the criticality region does not seem to be a really relevant parameter, contrary to what is expected in physics.

\end{document}